\documentclass[a4paper,fleqn]{cas-dc}
\usepackage[numbers]{natbib}
\usepackage[ruled]{algorithm2e}
\usepackage{subcaption}
\usepackage{flushend}
\usepackage{framed} 
\usepackage{multicol} 
 
\usepackage{nomencl} 
\makenomenclature
\renewcommand*\nompreamble{\begin{multicols}{2}}
\renewcommand*\nompostamble{\end{multicols}}

\usepackage{etoolbox}
\renewcommand\nomgroup[1]{%
  \item[\bfseries
  \ifstrequal{#1}{I}{Indices and Sets}{%
  \ifstrequal{#1}{P}{Parameters}{%
  \ifstrequal{#1}{V}{Variables}{}}}%
]}

\DeclareMathOperator*{\argmax}{arg\,max}
\SetKw{Continue}{continue}
\def\sectionautorefname{Section}
\def\algorithmautorefname{Algorithm}

\def\tsc#1{\csdef{#1}{\textsc{\lowercase{#1}}\xspace}}
\tsc{WGM}
\tsc{QE}
\tsc{EP}
\tsc{PMS}
\tsc{BEC}
\tsc{DE}

\begin{document}
\let\WriteBookmarks\relax
\def\floatpagepagefraction{1}
\def\textpagefraction{.001}
\shorttitle{Post disruption management for power system}
\shortauthors{Wu and Wang}

\title [mode = title]{Risk-Averse Optimization for Resilience Enhancement of Complex Engineering Systems under Uncertainties}                      
\author[1]{Jiaxin Wu}
\cormark[1]
\credit{Conceptualization of this study, Methodology, Software}
\address[1]{Industrial and Enterprise Systems Engineering, University of Illinois-Urbana Champaign, Urbana, IL, 61801, US}

\author[1]{Pingfeng Wang}
\cormark[2]
\credit{Conceptualization of this study, Methodology, Software}

\cortext[cor1]{Graduate Student}
\cortext[cor2]{Associate Professor, Corresponding Author}

\begin{abstract}
With the growth of complexity and extent, large scale interconnected network systems, e.g., transportation networks or infrastructure networks, become more vulnerable towards external disturbances. Hence, managing potential disruptive events during design, operating, and recovery phase of an engineered system therefore improving the system’s resilience is an important yet challenging task. In order to ensure system resilience after the occurrence of failure events, this study proposes a mixed integer linear programming (MILP) based restoration framework using heterogeneous dispatchable agents. Scenario based stochastic optimization (SO) technique is adopted to deal with the inherent uncertainties imposed on the recovery process from the nature. Moreover, different from conventional SO using deterministic equivalent formulations, additional risk measure is implemented for this study because of the temporal sparsity of the decision making in applications such as the recovery from extreme events. The resulting restoration framework involves with a large-scale MILP problem and thus an adequate decomposition technique, i.e. modified Langragian dual decomposition, is also employed in order to achieve tractable computational complexity. Case study results based on the IEEE 37-bus test feeder demonstrate the benefits of using the proposed framework for resilience improvement as well as the advantages of adopting SO formulations.
\end{abstract}




\maketitle

\section{Introduction}\label{sec:intro}
The increased complexity as well as more stringent requirements of reliability for interdependent critical infrastructures (ICIs) drive the research for a more robust and intelligent operational strategy. The goal is to analyze the performance of the components and subsystems in ICIs with the occurrence of different disruptive events during operation. Furthermore, cascading effects among the ICIs need to be mitigated therefore retaining their intended functionalities. Resilience, a term adopted from the ecology field, is used to measure how well a complex system can response towards disruptions and uncertainties \citep{DeAngelis1980}. Defining the system resilience benefits system operators and stakeholders in multiple aspects: (1) understanding how an ICI can autonomously detect and response to adversarial events under normal operating conditions and (2) how large the disruptions the system can withhold and (3) how quick the ICI can restore back to its nominal state \citep{Walker2004}. Different from the traditional view of system safety, i.e. minimizing the possibility of failures as well as the degraded system performance, system resilience has been considered as a complement to the safety metrics \citep{Steen2011, Francis2014, Righi2015}. Since a resilient system can not only resist and mitigate the negative effects of external or internal disruptive events, but also it can adapt to the disruptions and recover autonomously, based on the definitions from the report of the U.S. Department of Defense \citep{Goerger2014}.

In order to improve the system resilience and thus achieve system operations with better quality, current engineering resilience researches have been focusing on challenges in two temporal stages: before and after disruptions. For instance, researchers have proposed various different system modeling and analysis methods to quantify and analyze the resilience level of complex engineering systems, e.g. power distribution systems and supply chain networks, undergoing a disruptive event during the pre-disruption stage \cite{Yodo2016b,Sharma2017a}. Besides, approaches in the system design stage such as prognostic and health management, setting up redundancies (spare parts, backup, alternative), and repetitive maintenance (preventive or corrective) have been proposed to improve the resilience of the system \citep{Compare2014a, Youn2011a, Youn2011b, Wang2013}. Furthermore, probabilistic approaches such as Bayesian network has been adopted to analyze and quantify the overall system resilience with the presence of disruptions \citep{Yodo2016, Yodo2017}. However, one limitation of the methods focusing on the pre-disruptive stage is that they cannot guide how the system should perform after disturbances. As a result, severe disruptive events can lead to significant failures in the system and following cascading failures are eventual. Thus, in order to ensure a resilient system, additional measures to restore the system are needed. 

Motivated by the challenge of lacking a real-time system operational framework, studies have proposed different restoration frameworks for ICIs to recover from the disruptive state. As for the timely restoration after disruptions for networked systems, operational strategies, including but not limit to optimal repair scheduling \citep{Ouyang2017}, control guided recovery \citep{Wu2019, Yodo2018a}, and forming micro-grids \citep{Chen2016} are validated. In these works, the major task is to find the optimal decision for repair sequence or network reconfiguration actions throughout the entire recovery process in order to achieve the highest resilience level for the ICI. To extend the capability of the aforementioned decision making frameworks, \citet{Lei2019} propose a coordination framework combining the optimal scheduling of repair tasks and optimized routing of the distributed energy resources (DERs). In their work, the optimal decisions for the repair actions including locations and time steps as well as the dispatching plans for DERs can be determined by solving a mixed integer linear programming (MILP) problem. As a result, the power distribution systems considered in the case study can recover to the nominal state in the optimal sequence. Additionally, considering the limited budgets for recovery and the challenges during the initialization stage of the recovery, e.g. determining the appropriate capacity and resource level for repair crews (RCs) or DERs, \citet{Wu2020} present a coordination framework for recovery based on DERs and RCs coupled with a heuristic algorithm. In order to help the system achieve resilient response after disruptions, the proposed framework searches for the best practice in terms of recovery time and resource spent in a given time window. However, all aforementioned works are based on deterministic assumptions that the state awareness for the ICIs is omniscient: damaged locations or the magnitude of disruptions are predefined for the recovery framework. With the complex and extensive interactions between different subsystems as well as the highly uncertain recovery environment, uncertainties, e.g. uncertain damaged levels and corresponding required resources for repairing, can significantly deteriorate the performance of those recovery frameworks based on deterministic assumptions.

In order to tackle the challenges imposed by uncertainties in decision making process, techniques such as chance constrained optimization, robust optimization (RO) and stochastic optimization (SO) with recourse can be adopted. To list few relevant studies, \citet{Cao2013} study an optimal power flow management framework for the power distribution network by establishing a chance constraint optimization model considering uncertain generations from renewable energy resources. On the other hand, \citet{Gao2016} propose a chance constraint optimization model for critical load restoration by forming microgrids. In their study, uncertainties including intermittent energy generations and loads are included. As for RO approach, to better allocate the investment and preparation efforts for resilience, \citet{Fang2019c} evaluate the potential impacts from natural hazards on ICIs, for example, interconnected power-gas system, and treat the most-likely worst scenario as constraints in the RO framework for finding the best planning of ICIs. Moreover, \citet{Fang2019b} utilize conventional scenario based two-stage SO technique to find the optimal repair sequence for a power system after disruptions, while the required resource and the time for recovery is uncertain. It's noteworthy to point out that the optimal solution for chance constraint problem is often difficult to solve \citep{Ahmed2008}, and RO finds the optimal solution in terms of the worst realization of the scenarios with an uncertainty set \citep{Goh2011}. Thus, given its easiness to implement and profound ways for finding the optimal solution, the two-stage SO with recourse is used as the main decision making framework for heterogeneous agents to enhance system resilience. 

As for traditional two-stage SO with expected recourse function, it considers the expectation of the overall objective from numerous scenarios as the criterion to choose the optimal solution for random decisions \citep{Wilson1998}. And the result derived from the SO is risk neutral, which has large variance regarding to random outcomes. This risk neutral solution approaches the true optimum by repeatedly conducting decision makings under the similar condition. Such result is suitable in the long run for high frequency decision making problems, for instance, applications in the field of finance. However, recovery from extreme events to enhance system resilience is usually a non-repetitive decision making process and has significant temporal sparsity. Thus, acquiring a risk-averse solution that has smaller variability of the random outcomes after solving the SO model is more desirable for applications in enhancing system resilience. To accomplish the risk-averse requirement for SO solutions, additional risk measure can be incorporated into the framework, for example, the conditional value at risk (CVaR). 

The feasibility of utilizing CVaR in mathematical programming model to realize optimal decision making under uncertainties has been validated in numerous studies, for instance, the design of humanitarian relief network \citep{Noyan2012}, the supply chain management after disruptions \citep{Sawik2013}, the optimal facility location problem \citep{Yu2017a} and to optimize the battery storage cost in microgrids \citep{Tavakoli2018}. In this paper, we propose a MILP based post-disruption management (PODIM) framework using heterogeneous dispatchable agents, i.e. DERs and RCs, combined with the two-stage SO technique. The PODIM is formulated as a scenario based two-stage SO coupled with an additional risk measure CVaR. Due to the number of simulated scenarios for solving the SO problem, the resulting PODIM is a large-scale MILP problem with more than hundred thousand constraints and variables. As a result, an adequate decomposition technique, i.e. modified Lagrangian dual decomposition, is applied in order to achieve tractable solving time for the PODIM.

The rest of the paper is organized as follows: \sectionautorefname~\ref{sec:System_resilience} introduces the resilience metric used in this study as well as the definition of uncertainties considered in the PODIM. \sectionautorefname~\ref{sec:Model} demonstrates the mathematical formulations of the two-stage SO based PODIM framework coupled with the CVaR measure. \sectionautorefname~\ref{sec:Solving} discusses preprocessing steps and the decomposition technique to make the proposed recovery framework become computational tractable. A case study based on the IEEE 37-bus test feeder is presented in \sectionautorefname~\ref{sec:Results} in order to show the feasibility of the proposed PODIM, followed by \sectionautorefname~\ref{sec:Conclusion} where conclusion of this study is elaborated.

\section{System Resilience and Uncertainties}\label{sec:System_resilience}
This section introduces how to model the ICIs to incorporate any potential resilience enhancement technique as well as the concept of the system resilience. And the recovery framework has considered two uncertainties throughout the entire process: the required repair time for each damaged component and the associated resource level for repairing, determined by the uncertain repair task. Then the two-stage SO based recovery model integrated with additional risk measure is illustrated in detail.  

\subsection{Modeling of the ICIs and Resilience}
\begin{figure*}
    \centering
    \includegraphics[width = 0.8\textwidth]{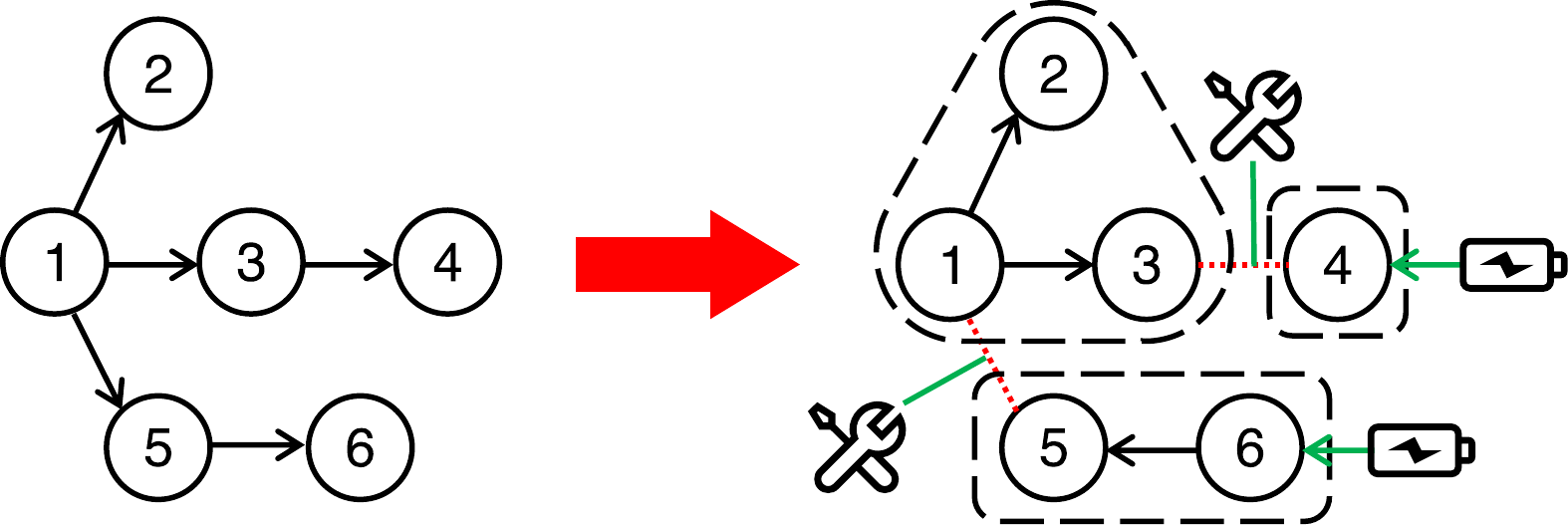}
    \caption{The radial graph representation of a networked system before and after disruptions with RCs and DERs}
    \label{fig:radial}
\end{figure*}

To address the interdependence and inherent networked structure of ICIs, the basic concept of graph theory is adopted in this study. For a general ICI, such as the commodity or electricity distribution system, the transportation hubs, warehouses, or buses can be modeled as nodes $\{i | i\in \mathbf{N}\}$. On the other hand, distribution paths are the edges $\{ij | ij\in \mathbf{L}\}$ between different nodes. Thus, the overall ICI can be denoted as a graph $\mathcal{G}:=(\mathbf{N},\mathbf{L})$. With a given disruption, various edges in the ICI will malfunction and become disconnected afterwards in the graph representation. By controlling remotely controllable switches on the intact edges, the original network can be reconfigured into several subsystems \citep{Carvalho2005,Moradi2008}. This study treats the ICI as a radial network, or a directed acyclic graph (DAG), which is represented as a tree in \figureautorefname~\ref{fig:radial}. This representation indicates that the flow on each edge $ij$ has a specific direction determined by the real-time operation conditions and no cyclical flow exists in the system. In such a way, the original system and the corresponding reconfiguration of the system after disruptions should form a single or several spanning trees as illustrated in \figureautorefname~\ref{fig:radial}, where red dot lines represent the damage edges. 

The motivation of modeling the system in radial topology is to simplify the problem formulation and avoid contradicted or cyclical commodity flow in the network. This radiality assumption is applicable for practical applications since most electricity distribution networks are designed to be radial in real life \citep{Murty2015,Peng2018}. To illustrate the idea, the original network shown in \figureautorefname~\ref{fig:radial} is a radial graph with $3$ branches, where node $1$ denotes the substation node of the network. After the disruption, edges $(1,5)$ and $(3,4)$ are disconnected, hence the network is partitioned into three sub-spanning trees, composed by node sets $\{1,2,3\}$, $\{4\}$, and $\{5,6\}$.

After a disruption, the network reconfiguration may isolates subsystems from substation nodes and disconnects the subsystems from the resource center. Thus those disconnected subsystems need to have effective backup resources in order to sustain the normal functionality after disruptions. This requirement motivates the first task of this study: the back-up resource should be deployed appropriately to support the normal operation of the ICI with the presence of disruptive events. One of such technique derived from deploying distributed backup sources for the power system is the microgrid formation \citep{Chen2016, Lei2019, Wu2020}. By forming microgrids, the original ICI can be divided into different partitions. Each partition is called one microgrid and is powered by dynamically allocated DERs, for example, mobile power generators as shown in \figureautorefname~\ref{fig:radial}. However, forming microgrids by using DERs is yet an emergency response procedure: ICIs remain in the damaged state with external backup resources. While the microgrids are formed, another task for decision makers is to dispatch repair crews with adequate repair tasks for damaged components to finally restore the damaged network to its nominal state. \figureautorefname~\ref{fig:radial} illustrates a restoration process that microgrids $\{4\}$ and $\{5,6\}$ are powered by DERs, while the damaged edges $(1,5)$ and $(3,4)$ are being repaired by RCs.

\begin{figure}
    \centering
    \includegraphics[width = 0.8\columnwidth]{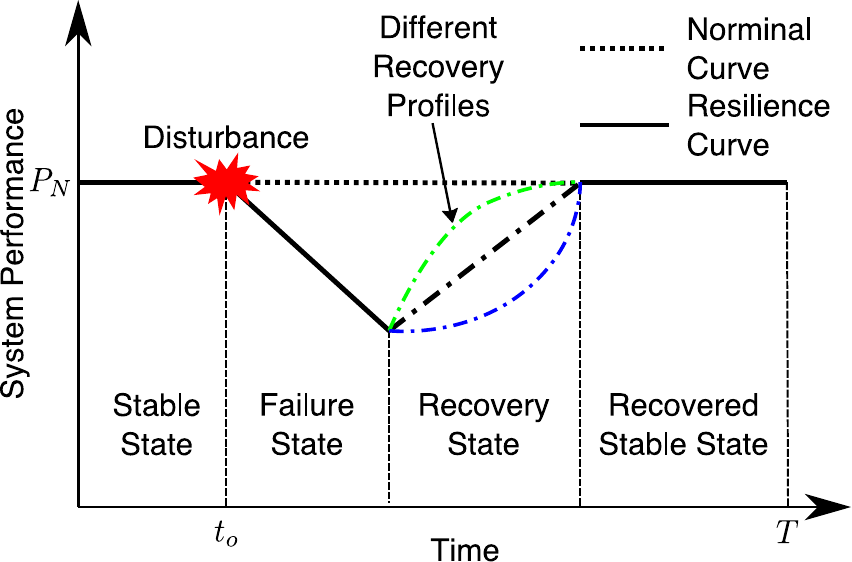}
    \caption{Resilience curve and 4 states in an ICI after the disruption}
    \label{fig:resilience}
\end{figure}

As a result, with the appropriate recovery framework, e.g. optimally assigning RCs and DERs to the network, the ICI can avoid the total failure even with the presence of disruptions and thus becomes resilient towards disturbance. To measure the system resilience, a typical resilience curve (real performance curve) with four states is illustrated in \figureautorefname~\ref{fig:resilience}. Note that the system performance curves could be different due to different resilience strategies during the recovery process.
Based on the resilience curve, this study defines the resilience level with respect to the system performance changes after the disturbance. By comparing the resilience curve with the nominal system performance curve, as shown in \figureautorefname~\ref{fig:resilience}, the resilience level can be measured. In the literature, several metrics for measuring resilience based on the resilience curve have been reported, and a review on the resilience metrics can be found in \cite{Yodo2016c}. In this study, the ratio of the area under the resilience curve to the area under the nominal performance curve is used as the resilience metric, which can be calculated mathematically as:
\begin{equation}\label{eqn:resilience}
    \Phi=\frac{\int_{t_0}^{T}RC(t)dt}{\int_{t_0}^{T}NC(t)dt},
\end{equation}
where $\Phi$ is the resilience level, $RC$ and $NC$ are the resilience curve and the nominal performance curve respectively, $t_0$ is the initial time before the occurrence of the disruption, while $T$ is the time that the recovery process has been finished and the system is considered to be settled at a new stable state. Intuitively this is true, since a larger area under the resilience curve generally means a smaller portion of the performance loss induced by the disruption, and thus the system is more resilient considering a given disruptive event.

\subsection{Modeling of Uncertainties}
As for recovering from failures for ICIs, the operational conditions have imposed numerous uncertainties, such as uncertain repair time or required resource, to worsen the performance of the recovery framework. In order to improve the applicability of the proposed recovery framework for practical applications, this study has considered two major uncertainties in the proposed framework: the uncertain repair time for different damaged components and the associated capacity of the required resource, e.g. manpower and equipment. 

\begin{figure}
    \centering
    \includegraphics[width = 0.8\columnwidth]{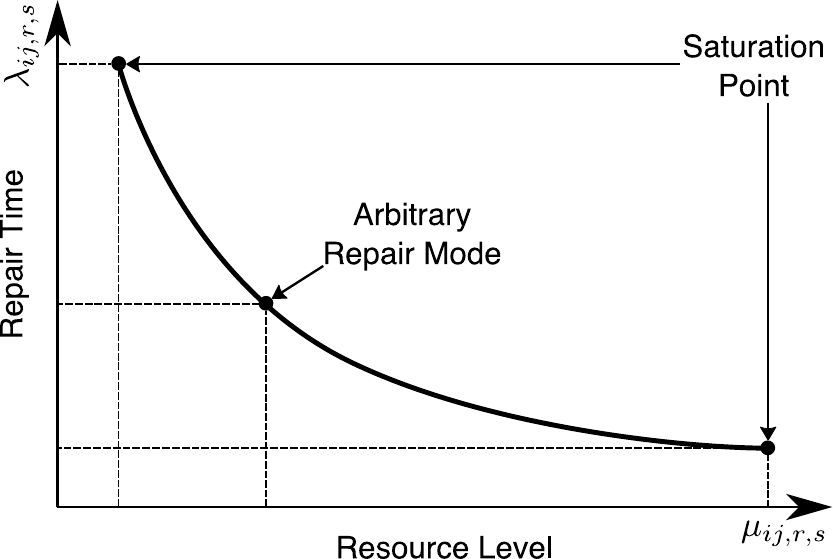}
    \caption{Repair time is decreasing with increased assigned resource but with diminishing returns}
    \label{fig:diminishing}
\end{figure}

Here, the well known two-parameters Weibull distribution from the field of reliability engineering is used to model the repair time for damaged components after disruptions \citep{Covert1973,Lai2003}. The mean of the Weibull distributed random variable, or the Mean Time to Repair (MTTR) is used as the parameter in the recovery model and can be derived from:
\begin{equation}\label{eqn:MTTR}
    MTTR=trepair_{ij,r,s}:=\lambda_{ij,r,s} \Gamma(1+k^{-1}),
\end{equation}
where $\lambda_{ij,r,s}$ is the scale parameter and $k$ is the shape parameter for the Weibull distribution. The other uncertainty, the required resource for repair, is treated as a normal distributed random variable $rs_{ij,r,s}\sim \mathcal{N}(\mu_{ij,r,s},\sigma^2)$ with expectation $\mu_{ij,r,s}$ and a fixed standard deviation $\sigma$. However, as for the repair tasks, it is intuitive to find that those two aforementioned uncertainties are not independent. For instance, assigning more repair resource or man power $rs_{ij,r,s}$ to an urgent task can effectively decrease the expectation of the uncertain repair time $trepair_{ij,r,s}$. Thus, \citet{Fang2019b} capture the relation between these two uncertainties by introducing a new variable: the repair mode $r$. The multimodal repair mode maps the required repair resource to the scale parameter of the MTTR following the law of diminishing returns, as shown in \figureautorefname~\ref{fig:diminishing}.
The convex curve demonstrates that assigning more resource can expedite one repair task but the resulting repair time is not decreasing linearly with increased resource accommodated. And the repair mode can be defined as any function $r:=\mathcal{F}(\mu_{ij,r,s})$, for instance the exponential function, that has the property of diminishing returns. On the other hand, this study sets two saturation points for the convex curve, that is, the repair time cannot be decreased/increased indefinitely by adding/removing the repair resource. This is consistent with the practical considerations that any repair task needs a small amount of resource to initialize and has a limited budget for restoration.

To incorporate the repair mode into the recovery framework for resilience as a MILP model, the convex function $\mathcal{F}(\mu_{ij,r,s})$ is discretized as a multimodal look-up table $(\mu_{ij,r,s},\\\lambda_{ij,r,s}),\;\forall ij\in\mathbf{L}$. Notice that in the PODIM, the repair mode is further determined by the binary decision variable $x_{ij}^r$ indicating whether damaged edge $ij$ is under repair mode $r$. As a result, the presented study in this paper aims to analyze the feasibility of improving system resilience by formulating optimal recovery process, specifically, the optimized scheduling of DERs and RCs as the control scheme after disruptions to ensure the resilience of ICIs in a real-time manner. Meanwhile, by formulating and solving a two-stage SO program, the system performance could be optimized with the presence of uncertainties during recovery. The scheduling of DERs or RCs has been well studied as optimization problems in the literature \citep{Ju2017,Li2017a}. In \sectionautorefname~\ref{sec:Model}, the mathematical formulations of a novel restoration framework by optimally coordinating DERs and RCs are introduced in detail.

\section{Two Stage SO based PODIM}\label{sec:Model}
In order to manage the ICIs after disruptions with the presence of uncertainties and thus achieve optimal resilience level, two different types of dispatchable agents are utilized to recover the system. This section discusses the formulations used by the two-stage SO model, which solves the optimal solutions of DERs and RCs scheduling.

\subsection{Two-Stage Stochastic Optimization Model}
Scenario based SO is a widely used technique to derive optimal solutions of MILP problems with uncertainties \citep{Shapiro2008}. In this study the two-stage SO formulation \citep{Pereira1991} is adopted to model the recovery framework for the ICI after disruptions and its generic determinnistic equivalent formulation is shown in \equationautorefname~\eqref{eqn:SO}:
\begin{equation}\label{eqn:SO}
    Z= \max \left\{cx+\sum_{j=1}^{S}p_j q_j y_j:x\in\mathbb{S}_x,\;y_j\in\mathbb{S}^j_y\right\},
\end{equation}
where $x$, $y_j$ are the first stage and second stage (recourse) decision variables correspondingly. The first stage $x$ is scenario independent while second stage decisions $y_j$ are scenario dependent. $p_j$ and $q_j$ are the endowed probability and cost of each simulated scenario $j$. $\mathbb{S}_x$ and $\mathbb{S}^j_y$ are the associated solution spaces for $x$ and $y_j$, defined as \equationautorefname~\ref{eqn:solution_space}:
\begin{equation}\label{eqn:solution_space}
    \begin{split}
    &\mathbb{S}_x:= \big\{ x: Ax \leq b, x\in X\big\},\\
    &\mathbb{S}^j_y:=\big\{T_j x+W_j y_j\leq h_j,y_j\in Y\big\},
    \end{split}
\end{equation}
in which corresponding constraints are brought into play. \equationautorefname~\ref{eqn:SO} indicates that, with the predefined probability of each scenario, the SO aims to maximize the first stage objective as well as the expectation of the second stage recourse problem. For this study, we treat finding the optimal repair mode $x_{ij}^r$ for each damaged component $ij$ in the ICIs as the first stage decision variable, while the second stage decisions $y_j$ involves with decisions for DERs and RCs.

\subsection{MILP Based Recovery Model}
In this section, the mathematical model including the objective function to maximize resilience and corresponding constraints for DERs/RCs scheduling are formulated. Constraints are categorized into four groups: DER dispatching constraints, RC scheduling constraints, network reconfiguration and system operation, as illustrated in \sectionautorefname~\ref{sec:DER} to \ref{sec:operation}.

\subsubsection{Objective for Enhancing Resilience}\label{sec:Objective}
During the operation of restoring power distribution systems by dispatching DERs and RCs, loads have different priorities to be picked up, according to their importance to sustain critical functions in the ICI. Thus, for the load restoration after the blackout due to a disruption, those important loads should be given higher priorities. Let $w_i$ denotes the priority weight associated with the load at node $i$, and a larger value of weight indicates a higher priority. Then, the overall objective of the PODIM framework following the generic two-stage SO objective in \equationautorefname~\ref{eqn:SO} can be expressed as:
\begin{equation}\label{eqn:vanilla_objective}
    \max\mathbb{E}\left[\sum_{t\in\mathbf{T}}\sum_{i\in\mathbf{N}}w_i \cdot o_{i,t,s} \cdot p_{i,t}\right].
\end{equation}
This objective of the PODIM aims to maximize the load picked up after disturbances throughout the entire recovery process. And the resulting resilience curve $RC(t)$ will have greater area under the curve as shown in the resilience plot in \figureautorefname~\ref{fig:resilience}. Notice that different from \equationautorefname~\ref{eqn:SO}, this object only includes the cost for the second stage recourse problem since the first stage decision $x_{ij}^r$ involves with no cost. And with the maximized load picked up, the resilience level in terms of system performance, i.e. satisfied load in the network after disruptions, can be optimized.

\subsubsection{Constraints for DER Assignment}\label{sec:DER}
After severe disruptions, the power distribution network may be isolated from the main grid due to damaged lines. With appropriate design and operational planning, the restoration process starts immediately after the damage has been assessed. In this study, DERs are dispatched and routed dynamically to energize the power distribution network along with repair scheduling after the disruption happens. The PODIM framework is focusing on using mobile generators as the backup resource and formulating the dispatching as an optimization problem in order to maximize the restored load. Followings are the corresponding constraints for the DER dispatching problem.
\begin{gather}
    \sum_{i \in {{\bf{N}}_{\bf{G}}}} {\alpha _{i,t,s}^g \le 1,} \forall g \in {\bf{G}}, \forall t \in {\bf{T}}, \forall s \in {\bf{S}}, \label{eqn:model_eqn_1} \\
    \sum_{g \in {\bf{G}}} {\alpha _{i,t,s}^g \le 1,} \forall i \in {{\bf{N}}_{\bf{G}}}, \forall t \in {\bf{T}}, \forall s \in {\bf{S}}, \label{eqn:model_eqn_2}\\
    \begin{split}\label{eqn:model_eqn_3}
        \alpha _{i,t,s}^g + \alpha _{j,t + \tau,s}^g \le 1, \forall g \in {\bf{G}}, &\forall i,j \in {{\bf{N}}_{\bf{G}}}, \forall s \in {\bf{S}},\\
        &\forall \tau \le tr_{ij}^g, \forall t + \tau  \le \bf{T},
    \end{split}\\
    0 \le Pg_{t,s}^g \le \sum_{i \in {{\bf{N}}_{\bf{G}}}} {\alpha _{i,t,s}^g \cdot {{\overline {Pg} }_g}}, \forall g \in {\bf{G}}, \forall t \in {\bf{T}}, \forall s \in {\bf{S}}. \label{eqn:model_eqn_4}
\end{gather}
Constraints \ref{eqn:model_eqn_1} to \ref{eqn:model_eqn_4} determines feasible routines of the DER for recovery. Both the number of nodes that can be energized by the same DER and the number of DERs that can be connected to the same node are set as 1 by implementing constraints \ref{eqn:model_eqn_1} and \ref{eqn:model_eqn_2}. Constraint \ref{eqn:model_eqn_3} defines the status of traveling for the DER, which implies that the same DER $g$ cannot be at a different node until the traveling time $tr_{ij}^g$ has been bypassed. In constraint \ref{eqn:model_eqn_4}, the power generation from each DER is governed by both the maximum capacity $\overline{Pg}_g$ and the working status $\alpha_{i,t,s}^g$ of the DER.

\subsubsection{Constraints for RC Scheduling}\label{sec:RC}
In the previous section, how to coordinate DERs to energize the damaged ICI and partially restore the functionality is introduced. However, scheduling repair tasks is needed to fully restore the damaged components and to bring the system back to its nominal state. In this study, repair tasks are accomplished by several RCs. And their appropriate schedules are formulated by following constraints, adopted from \cite{Fang2019b}.
\begin{gather}
    \sum_{r \in {\bf{R}}} {x_{ij}^r \le 1,} \forall ij \in {\bf{L}}_{t}^{{\bf{damage}}},\label{eqn:model_eqn_5}\\
    \sum_{t=1}^{T}\phi_{ij,t,s}=\sum_r x_{ij}^r, \forall ij \in {\bf{L}}_{t}^{{\bf{damage}}}, \forall s \in {\bf{S}},\label{eqn:model_eqn_6}\\
    \begin{split}\label{eqn:model_eqn_7}
        a_{ij,t,s}+\mu_{ij,t,s}=\sum_{\tau=t_0}^{t}&\phi_{ij,\tau,s},\\
        &\forall ij \in {\bf{L}}_{t}^{{\bf{damage}}},\forall t \in \bf{T}, \forall s \in {\bf{S}},    
    \end{split}\\
    \mu_{ij,t,s}\leq \mu_{ij,t+1,s}, \forall ij \in {\bf{L}}_{t}^{{\bf{damage}}}, \forall s \in {\bf{S}}, \label{eqn:model_eqn_8}\\
    \begin{split}\label{eqn:model_eqn_9}
        \sum_{t}t\phi_{ij,t,s}+\sum_{t}a_{ij,t,s}&\leq \sum_{t}t[\mu_{ij,t,s}-\mu_{ij,t-1,s}], \\ 
        &\forall ij \in {\bf{L}}_{t}^{{\bf{damage}}}, \forall t \in \bf{T}, \forall s \in {\bf{S}}, 
    \end{split}\\
    \begin{split}\label{eqn:model_eqn_10}
    \sum_{t} a_{ij,t,s} \geq \sum_r & trepair_{ij,r,s}\cdot x_{ij}^r,\\ 
    &\forall ij \in {\bf{L}}_{t}^{{\bf{damage}}}, \forall t \in \bf{T}, \forall s \in {\bf{S}}, 
    \end{split}\\
    \sum_{ij}(a_{ij,t,s}\sum_r rs_{ij,r,s}x_{ij}^r)\leq \overline {RS}_{t,s}, \forall t \in \mathbf{T}, \forall s\in\mathbf{S}. \label{eqn:resource}
\end{gather}
Constraint \ref{eqn:model_eqn_5} states that only one repair mode should be chosen for each damaged component after disruptions. The repair mode is represented as the binary first stage decision variable $x_{ij}^r$ indicating whether the damaged edge $ij$ is repaired under mode $r$. And constraint \ref{eqn:model_eqn_6} indicates that resources have to be assigned if any damaged component is selected to be repaired under any mode. Once the required repair time has been bypassed and one damaged component has resource as well as repair actions assigned throughout the entire time span, then the associated operational status $\mu_{ij,t,s}$ should be updated, as shown in Equation \ref{eqn:model_eqn_9} and \ref{eqn:model_eqn_10}. Notice here the MTTR is used as the required repair time $trepair_{ij,r,s}$ and is assumed to be a Weibull distributed random variable. And the mean of the random variable is based on the repair mode assigned: larger resource level $rs_{ij,r,s}$ assigned to one repair task leads to shorter required repair time, i.e. uncertain MTTR with explicit smaller mean. Besides random repair time, another uncertainty $rs_{ij,r,s}$ is arisen in constraint \ref{eqn:resource}, where the total resource assigned to conduct the repair tasks has to be smaller than the available resources $\overline{RS}_{t,s}$ the decision maker can dispatch.

\subsubsection{Constraints for Radiality of the Network}\label{sec:radiality}
After a disruption, the power distribution system can avoid entire blackout by forming multiple microgrids with the help of remotely controllable switches and DERs. In many researches, the microgrids formed are treated as several spanning trees or radial graphs \citep{Chen2016, Lei2019}. This study adopts the same manner and makes the original distribution network form radial topology after disruptions with disconnected distribution lines. According to \citet{Lavorato2012a}, constraints \ref{eqn:model_eqn_12} to \ref{eqn:model_eqn_17} are sufficient to ensure the radiality of the reconfigured distribution network. 
\begin{gather}
    {\mu _{ij,t,s}} = 1, \forall ij \in {\bf{L}}\backslash {\bf{L}}_{\bf{t}}^{{\bf{damage}}}, \forall s \in {\bf{S}}, \label{eqn:model_eqn_12}\\
     {\mu _{ij,0,s}} = 0, \forall ij \in {\bf{L}}_{\bf{t}}^{{\bf{damage}}}, \forall s \in {\bf{S}}, \label{eqn:model_eqn_13}\\
     {\beta _{ij,t,s}} \le {\mu _{ij,t,s}}, \forall ij \in {\bf{L}},\forall t \in {\bf{T}}, \forall s \in {\bf{S}},	\label{eqn:model_eqn_14}\\
     \sum_{ij \in {\bf{L}}} {{\varepsilon _{ij,t}} = N - 1}, \forall t \in {\bf{T}}, \label{eqn:model_eqn_15}\\
     \sum_{ji \in {\bf{L}}} {{f_{ji,t}}}  - \sum_{ij \in {\bf{L}}} {{f_{ij,t}} = 1}, \forall t \in {\bf{T}},	\label{eqn:model_eqn_16}\\
     {\beta _{ij,t,s}} \le {\varepsilon _{ij,t}}, \forall ij \in {\bf{L}},\forall t \in \bf{T}, \forall s \in {\bf{S}}. \label{eqn:model_eqn_17}
\end{gather}
Constraints \ref{eqn:model_eqn_12} and \ref{eqn:model_eqn_13} initialize the operational status for both damaged and intact edges. And constraint \ref{eqn:model_eqn_14} says that edges in the network can only be connected (ON) if they are operational. Furthermore, in order to guarantee the radiality of the network, a fictitious undamaged network with the same topology as the original real system is employed. In the radial, fictious network, constraint \ref{eqn:model_eqn_15} states that the number of connected lines equals the total number of nodes minus one. And constraint \ref{eqn:model_eqn_16} indicates the fictitious flow should present on all edges in the undamaged network. Once the fictious network is assured to be radial, the radiality of the real system can be guaranteed by superimposing the connection status variables $\epsilon_{ij,t}$ to $\beta_{ij,t,s}$, as shown in \equationautorefname~\ref{eqn:model_eqn_17}.

\subsubsection{Constraints for System Operation}\label{sec:operation}
Besides constraints for assigning RCs/DERs and the topology of the network, some physical constraints such as power flow balance need to be fulfilled, as shown in constraints \ref{eqn:model_eqn_18} to \ref{eqn:model_eqn_last}. For instance, constraint \ref{eqn:model_eqn_18} formulates the flow balance between different nodes by using the Kirchhoff's law. The power flow has its maximum value restricted by \ref{eqn:model_eqn_19}. Constraints \ref{eqn:model_eqn_20} and \ref{eqn:model_eqn_21} show that the power generation from DERs is coupled with the DER assignment variable. And the power flow from the fixed substation node is regulated by the maximum capacity $\overline{G}_{i,t}$ based on constraint \ref{eqn:model_eqn_22}. Constraint \ref{eqn:model_eqn_last} assumes that once a load has been restored, it cannot be dropped again during the recovery process.
\begin{gather}
    \begin{split}\label{eqn:model_eqn_18}
        \sum_{ji \in {\bf{L}}} {{P_{ji,t,s}}}  + P{g_{i,t,s}} + {G_{i,t,s}} &= \sum_{ij \in {\bf{L}}} {{P_{ij,t,s}} + {o_{i,t,s}} \cdot {p_{i,t}}}, \\
        &\forall i,j \in {\bf{N}}, \forall t \in {\bf{T}}, \forall s \in {\bf{S}},
    \end{split}\\
     {P_{ij,t,s}} \le {\beta _{ij,t,s}} \cdot {\overline {flow} _{ij}}, \forall ij \in {\bf{L}}, \forall t \in {\bf{T}}, \forall s \in {\bf{S}},\label{eqn:model_eqn_19}\\
     P{g_{i,t,s}} = \sum_{g \in {\bf{G}}} {\alpha _{i,t}^g \cdot Pg_{t,s}^g}, \forall i \in {{\bf{N}}_{\bf{G}}}, \forall t \in {\bf{T}}, \forall s \in {\bf{S}}, \label{eqn:model_eqn_20}\\
     P{g_{i,t,s}} = 0, \forall i \in {\bf{N}}\backslash {{\bf{N}}_{\bf{G}}}, \forall t \in {\bf{T}}, \forall s \in {\bf{S}}, \label{eqn:model_eqn_21}\\
     0 \le {G_{i,t,s}} \le {\overline{G}_{i,t}},\; \forall i \in {\bf{N}}, \forall t \in {\bf{T}}, \forall s \in {\bf{S}}, \label{eqn:model_eqn_22}\\
     {o_{i,t,s}} \le {o_{i,t + 1,s}}, \forall i \in {\bf{N}}, \forall t \in {\bf{T}}, \forall s \in {\bf{S}}. \label{eqn:model_eqn_last}
\end{gather}

\subsection{Risk-averse optimization Objective}
Recall in the previous section, the objective of the two-stage SO based PODIM is to maximize the system resilience among different sampled scenarios, as shown in \equationautorefname~\ref{eqn:vanilla_objective}. However, one drawback of the solutions derived from the conventional scenario based SO is that they may come with large variance. As for the decision-making process with high frequency, the large variance can be overcame by repeatedly making decisions following the SO solution: the true outcome will be closed to the expectation of the scenario based stochastic solutions because of the large number theory \citep{Noyan2012}. Nonetheless, temporal sparsity in the decision-making process for rare events, e.g. recovery from disruptions for power systems, prevents the real outcome from following the expectation exactly with a large number of trails. As a result, risk-averse optimization can be adopted here to decrease the variance in the solutions in order to achieve more confident recovery performance. And the CVaR is employed as an additional risk measure in the objective function in this study. To realize the risk-averse optimization framework, two additional constraints for calculating the risk measure are added to the PODIM:
\begin{gather}
      Restoration_s = -cost_s=\sum_{t \in {\bf{T}}} {\sum_{i \in {\bf{N}}} {{w_i}}  \cdot {o_{i,t,s}} \cdot {p_{i,t}}} \label{eqn:CVaR_1},\\
     -cost_{s}+\nu\geq -\Delta_{s}, \forall s \in \mathbf{S} \label{eqn:CVaR_2},
\end{gather}
where $\nu$ is the value at risk (VaR) of the solutions for all scenarios and $\Delta_s$ is the excessive amount above the VaR, for all solutions that are greater than the VaR. In order to compute the CVaR measure, $\nu$ is considered as a newly introduced first stage decision variable. On the other hand, the scenario dependent variable $\Delta_s$ is treated as a new second stage variable, in parallel with the decision variables for DERs and RCs. To summarize, the new risk-averse objective of the PODIM coupled with the CVaR criterion becomes:
\begin{equation}\label{eqn:Objective}
    \begin{gathered}
        \min \lambda(\nu+\frac{1}{1-\alpha}\frac{1}{n}\sum_{s=1}^{n}\Delta_{s})-\mathbb{E}(\sum_{t \in {\bf{T}}} {\sum_{i \in {\bf{N}}} {{w_i}}  \cdot {o_{i,t,s}} \cdot {p_{i,t}}}),\\
        w.r.t\;(\ref{eqn:model_eqn_1})-(\ref{eqn:model_eqn_last}),\;(\ref{eqn:CVaR_1})\;(\ref{eqn:CVaR_2})
    \end{gathered}
\end{equation}
where two additional hyperparameters $\lambda$ and $\alpha$ are added to control the derived risk-averse solution. Here, $\lambda$ is the weight for the additional CVaR measure, while $\alpha$ is the significance level for solving the VaR value $\nu$. Different $\alpha$ will lead to different VaR result. Due to the limited space, the proof of the derivation for the CVaR objective is omitted in this paper. Interested readers can refer to \citep{Rockafellar2000, Rockafellar2002} for more in-depth understanding.

\section{Solving Process}\label{sec:Solving}
In order to efficiently solve the MILP model proposed in \sectionautorefname~\ref{sec:Model}, several steps need be taken in account. First, nonlinear constraints should be linearized before solving the model by any commercial linear programming solver. Moreover, since the DERs are dynamically routed during the recovery process in the ICIs, determining the initial candidate nodes for allocating the DERs is important for ensuring robust performance. Furthermore, two-stage SO model usually involves with decomposition steps to tackle the challenge of high computational complexity, induced by the size of the problem. Therefore, this section discusses required preprocessings and the solving method for the PODIM.

\subsection{Linearization}
In the PODIM model proposed in \sectionautorefname~\ref{sec:Model}, \equationautorefname~\ref{eqn:resource} is a nonlinear constraint. Since a decision variable $a_{ij,t,s}$ multiplies with another integer variable $x_{ij}^r$, and forms a quadratic term in \equationautorefname~\ref{eqn:resource}. Standard linear programming solvers cannot easily solve this type of constraints. Hence, Big-M method is adopted to linearize the nonlinear term for this study. Here two auxiliary variables $w_{ij,s}$ and $z_{ij,t,s}$ are first defined as below:
\begin{gather}
    w_{ij,s} = \sum_r {rs_{ij,r,s}}x_{ij}^r, \label{eqn:linearize_1}\\
    z_{ij,t,s} = a_{ij,t,s}w_{ij,s} \label{eqn:linearize_2}.
\end{gather}
Nevertheless, the new auxiliary variable $z_{ij,t,s}$ is also a quadratic term that requires additional steps to derive the final linear result. Since the decision variable $x_{ij}^r$ is binary, the upper bound of $w_{ij,s}$ can be expressed as:
\begin{equation}
    w_{ij,s}\leq\overline{w}_{ij,s}=\sum_r rs_{ij,r,s}, \forall ij\in\mathbf{L}_t^{\text{damage}}, \forall s\in\mathbf{S}\label{eqn:linearize_3}.
\end{equation}
Based on this upper bound, two inequalities can be used to fix the value of $z_{ij,t,s}$ rather than defining this auxiliary variable by multiplying two decision variables:
\begin{gather}
    0\leq z_{ij,t,s} \leq a_{ij,t,s}\overline{w}_{ij,s},\\
    w_{ij,s}-(1-a_{ij,t,s})\overline{w}_{ij,s}\leq z_{ij,t,s} \leq w_{ij,s}.
\end{gather}
Following the two given inequalities, when the binary variable $a_{ij,t,s}$ is 0, the first inequality is active and set $z_{ij,t,s}$ to be 0. On the other hand, the second bound is active when $a_{ij,t,s}$ equals 1 and $z_{ij,t,s}$ is $w_{ij,s}$. And this result is consistent with the expression in \equationautorefname~\ref{eqn:linearize_2}. In terms of the auxiliary variables as well as the bounds introduced, \equationautorefname~\ref{eqn:resource} can be rewritten as:
\begin{equation}
    \sum_{ij} z_{ij,t,s}\leq \overline{RS}_{t,s}\;\forall t\in\mathbf{T},\;\forall s\in\mathbf{S}. \label{eqn:linearize_6}
\end{equation}
Consequently, the nonlinear constraint is reformulated by three auxiliary variables as well as a combination of four linear constraints defined in \equationautorefname~\ref{eqn:linearize_3}-\ref{eqn:linearize_6}. The overall PODIM model becomes a mathematical programming problem with only linear constraints.

\subsection{Preassign DERs}
For DERs to energize the disconnected subsystems after disruptions, decision makers need to first determine the connecting points for DER assignments. Instead of randomly initializing the starting positions for DERs and waiting for the MILP model to search for the optimal locations, we can predefine the most suitable candidate positions for DERs by graph algorithms to achieve better performance in practical applications. Given an assessment of malfunctioning components after disturbances in the ICIs, the disconnected subgraphs can be identified by using graph search algorithms, for example, breadth-first-search or depth-first-search. After the subgraphs have been recognized, the system operator can preassign DERs to those isolated regions even before starting the repair tasks. Because of the radiality of the original networked system, each subgraph is also a tree. Therefore, different locations to host the DER lead to the same transmission cost for the power flow in one subgraph. However, different candidate nodes to connect the DER can be differentiated by the centrality of the allocated DER. 

In order to enhance the robustness of the deployed DER, decisions for the DER location prefer higher centrality. And in this study the degree centrality, i.e. the number of neighbors incident upon the node, is employed to evaluate the quality of the DER location. The overall algorithm to preassign the DERs based on centrality for each subgraph after disruptions is summarized in \algorithmautorefname~\ref{Alg:DER}. $V$ is a set representing whether the vertex $v$ in the graph has been explored by BFS; $T$ is a 2D array storing the vertices for each disconnected subgraph $i$; $D$ is the map recording the degree of each node in a subgraph; $n$ is the total number of disconnected subgraphs after disruptions.
\begin{algorithm}[!t]
\SetKwInOut{Input}{Input}
\SetKwInOut{Output}{Output}
\SetAlgoLined
\Input{Adjacency matrix $A$ of the system after disruptions.}
\Output{The initial nodes $\mathbf{N_G}$ to host DER for disconnected subgraphs after disruption.}
$V\gets \{\;\}$, $T\gets [\;]$, $n\gets 0$\;
\BlankLine
\While{$\vert V \vert <$ number of nodes}{
    $S\gets$ the first vertex with status $0$ in $V$\;
    Run BFS on $A$ with $S$ as the starting vertex\;
    $V\gets$ explored vertices\;
    $T\gets$ a connected component identified by BFS\;
    $n\gets n+1$\;
}
\For{$i<n$}{
    $D\gets \{\;\}$\;
    \For{$v\in T[i]$}{
        $D[v]\gets \text{number of neighbors of $v$}$\;
    }
    \eIf{substation node $\notin T[i]$}{
    $\mathbf{N_G}[i]=\argmax{D[v]}$\;
    }{
    \Continue\;
    }
}
\caption{Preassign DERs}\label{Alg:DER}
\end{algorithm}

\subsection{Decomposition}
One significant drawback of the the scenario based SO is the large computational complexity due to the need of sampling numerous scenarios. For example, sampling 50 scenarios for the proposed PODIM makes the problem have more than 1000 continuous variables and 100000 integer variables with 50000 constraints. Moreover, with the presence of complex variables, i.e. the first stage decision variables, the problem requires significant efforts to solve. To enhance the computational tractability, decomposition techniques, such as the Bender’s decomposition, are widely used for SO problems \citep{Geoffrion1972,Geoffrion1974,Binato2001}. Traditional Bender's decomposition relies on the assumption that second stage of the model only involves with continuous linear or nonlinear program. To be specific, Bender's decomposition first treats the integer variables at the first stage as known constants. And the original problem $f(x,y)$ can be decomposed into a relaxed master problem $f(x)$ with multiple sub-problems $Q_j(\bar{x},y_j)$, which only have continuous variables $y_j$ as well as constants $\bar{x}$. Based on the solutions of the linear sub-problems, optimality and feasibility cuts (constraints) can be generated and added to the master problem because of the duality. With the generated solution cuts, the master problem and the subproblems can be solved iteratively to approach the actual optimum $x^*$ and $f^*$. 

The proposed PODIM relies on dispatching DERs and RCs in the second stage, which can be highly combinatorial and thus the second stage program usually has integer decision variables. And these integer variables prevent the overall model from being decomposed into subproblems with integer or continuous variables solely. In this case, Bender's decomposition can not be used as a solving schema for the PODIM. Yet, we examine the feasibility of applying the Lagrangian decomposition for SO to decrease the computational complexity. \citet{Carhe1999} first proposed this algorithm to solve stochastic integer programming problems with recourse. Later \citet{Schultz2006} extended the algorithm to risk-averse optimization with CVaR as the risk measure. To be self-contained, here we briefly summarize the derivation of the algorithm and the corresponding modifications in order to apply the decomposition to the proposed PODIM.

The overall objective of the PODIM shown in \equationautorefname~\ref{eqn:Objective} can be reformulated into the form of a general two-stage SO with recourse problem:
\begin{equation}\label{eqn:Decomposition_1}
    \begin{split}
            &Z^*=\min cx+b\nu+\sum_{j=1}^S p_jq^T_jy_j, \\
            &s.t.\;x\in \mathbb{S}_x,\;\nu\in \mathbb{S}_\nu,\;y_j\in \mathbb{S}^j_y.
    \end{split}
\end{equation}
The $x$ and $\nu$ here are the first stage scenario independent variables; $y_j$ is the second stage decision variable; $\mathbb{S}_x$, $\mathbb{S}_\nu$ and $\mathbb{S}_y^j$ are the solution spaces determined by the constraints for first and second stage variables. Notice that for the PODIM, $x$ is the repair mode $x_{ij}^r$ for each damaged component, $\nu$ is the VaR for the risk measure derivation, and $y_j$ is the operational decision variable in each scenario.

The first stage variables are scenario independent and have to be not only feasible but also optimal for all uncertain scenarios. This characteristic makes $x$ and $\nu$ be complex variables. In order to relax the consistence of $x$ and $\nu$ crossing all scenarios, the variable splitting technique can be utilized \citep{Afonso2010}. After adding $\vert S \vert-1$ copies of the scenario independent variables. i.e. $x$ and $\nu$, to the program, \equationautorefname~\ref{eqn:Decomposition_1} can be rewritten as:
\begin{equation}\label{eqn:Decomposition_2}
    \begin{split}
        &Z^*=\min\sum_{j=1}^S p_j(c_jx_j+b_j\nu_j+q^T_jy_j),\\
        &s.t.\;x\in \mathbb{S}^j_x,\; \nu_j\in\mathbb{S}^j_\nu,\; y_j\in \mathbb{S}^j_y,\\
        &x_1=\hdots=x_s,\;\nu_1=\hdots=\nu_s,\;\forall j=1,\hdots,S
    \end{split}
\end{equation}
In order to enforce the consistence among the additional copies of the first stage decision variables, the nonanticipativity constraints $x_1=\hdots=x_s$ and $\nu_1=\hdots=\nu_s$ are incorporated. Since $x$ is a binary variable in our case, instead of adding $\vert S \vert-1$ new equality constraints, the nonanticipativity constraint for $x$ can be rewritten into a compound form :
\begin{equation}\label{eqn:Decomposition_3}
    \left(\sum_{j=2}^S c_j\right)x_1=c_2x_2+\hdots+c_Sx_S.
\end{equation}
Consequently, after splitting the scenario independent variables, the original MILP problem changes from a program with coupled variables to a mathematical programming problem with complex constraints, i.e. the additional nonanticipativity constraints. Hence appropriate constraint relaxation can be adopted to decompose the program. Take Lagrangian relaxation with respect to the nonanticipativity constraint, the problem becomes:
\begin{equation}\label{eqn:Decomposition_4}
    \begin{split}
        &L(x_j,\nu_j,y_j,\lambda)=\min\sum_{j=1}^Sp_j(c_jx_j+b_j\nu_j+q^T_jy_j)\\
        &\qquad\qquad\qquad\qquad\quad-\lambda_1\sum_{j=1}^S H_j x_j-\lambda_2\sum_{j=1}^S K_j \nu_j\\
        &s.t.\;x\in \mathbb{S}^j_x,\;\nu\in \mathbb{S}^j_\nu\;y_j\in\mathbb{S}^j_y,\;\forall j=1,\hdots,S.
    \end{split}
\end{equation}
Here, $\lambda_1$ and $\lambda_2$ are the Lagrangian multipliers, and are different from the $\lambda$ in \equationautorefname~\ref{eqn:Objective}. $H$ and $K$ are the parameter matricies for the representation of nonanticipativity constraints, for instance, the compound representation for $x$ as shown in \equationautorefname~\ref{eqn:Decomposition_3}. The Lagrangian dual of \equationautorefname~\ref{eqn:Decomposition_4} can be derived as:
\begin{equation}\label{eqn:Decomposition_5}
    Z^{LD}=\max_{\lambda}\min_{x_j,\nu_j,y_j}L(x_j,\nu_j,y_j,\lambda),
\end{equation}
Based on the weak duality and since $L(x_j,\nu_j,y_j,\lambda)$ is a relaxed solution for $Z^*$, we know that the objective value $Z^{LD}$ provides an lower bound for the actual optimality of minimizing $Z^*$, i.e. $Z^{LD} \leq Z^*$. The derived dual program in \equationautorefname~\ref{eqn:Decomposition_5} usually is a non-smooth concave maximization problem. In order to solve it, subgradient method can be used \citep{Sherali1996}. For instance, the subgradient of \equationautorefname~\ref{eqn:Decomposition_4} can be calculated as $\left(\sum_{j=1}^S H_jx_j,\sum_{j=1}^S K_j\nu_j\right)\in\partial L(\lambda)$.

The obtained dual program $Z^{LD}$ is a linear program without any complex variable or constraint and is much easier to solve than the original problem as shown in \equationautorefname~\ref{eqn:Decomposition_1}. Moreover, the dual program is a linear combination of $\vert S \vert$ problems that are sharing the identical structure. As a result, the solution $Z_j^{LD}$ from optimizing the Lagrangian dual program of each scenario can replace the linear relaxation solution in the traditional branch-and-cut algorithm for solving large scale MILP problems. And since we solve the dual program of the original problem, the derived solution provides a bound for the true optimum. With the more sophisticated bound, the MILP based PODIM can be solved to optimum much faster than the case of using the vanilla branch-and-bound technique. \algorithmautorefname~\ref{Alg:MIP_BB} summarizes the modified branch-and-bound algorithm with the Lagrangian dual decomposition. In the algorithm, $Q(x,\nu)$ represents the second stage scheduling problem based on the solutions of repair mode and VaR. Once all copies of first stage variables are identical and subproblems of all scenarios have been examined, we know that the optimum has been found. 

The advantage of the branch-and-bound with dual decomposition (DD) is that it exploits the similar structures crossing different scenarios of the SO problem. And incorporated variable splitting technique allows slightly inconsistency of the first stage decision variables by creating copies among different scenarios and shrinks the discrepancy by branching on, as well as bounding the suboptimal solutions. In the contrast, the traditional branch-and-bound algorithm with linear relaxation treats all scenarios as a large scale all-in-one problem. Furthermore, the branch and cut algorithm used by the commercial solver, e.g. CPLEX and Gurobi, handles the consistence of the first stage variables as a hard constraint and thus tries to approach the optimum by solving a bulk of problems simultaneously. To conclude, programs solved by the DD can converge to the optimum faster but have infeasible solutions at the initial time steps, while traditional branch-and-bound algorithm starts with a feasible solution but requires much longer convergence time.
\begin{algorithm}[t]
\SetAlgoLined
\SetKwInOut{Input}{Input}
\SetKwInOut{Output}{Output}
\SetKw{Break}{break}
\SetKw{Continue}{continue}
\Input{SO formulation in \equationautorefname~\ref{eqn:Decomposition_2}}
\Output{Bounded near optimal solution}
$Z\gets+\infty,\mathcal{P}\gets \{Z^*\,\text{of each scenario}\}$\;
\BlankLine
\While{$\mathcal{P}\neq\emptyset$}{
    Select and delete a problem $P$ from $\mathcal{P}$\;
    Solve $Z^{LD}(P)$ of $P$ as shown in \equationautorefname~\ref{eqn:Decomposition_5}\;
    \uIf{$P$ is infeasible}{
    \Continue\;
    }\uElseIf{$Z^{LD}(P)\geq Z$}{
    \Continue\;
    }\uElseIf{$x_j$'s are identical and $\nu_j$'s are identical}{
    $x\gets x_j$,\,$\nu\gets \nu_j$\;
    $Z=\min\{Z,cx+b\nu+\mathbb{E}[Q(x,\nu)]\}$\;
    delete all P with $Z^{LD}\geq Z$ from $\mathcal{P}$\;
    \Continue\;
    }\ElseIf{$x_j$'s are different or $\nu_j$'s are different}{
    compute average $\bar{x}$, $\bar{\nu}$\;
    round $\bar{x}$ to a binary $\bar{x}^R$\;
    $Z=\min\{Z,c\bar{x}^R+b\bar{\nu}+\mathbb{E}[Q(\bar{x}^R,\bar{\nu})]\}$\;
    delete all P with $Z^{LD}\geq Z$ from $\mathcal{P}$\;
    select a component $x_i$ of all $x_j$\;
    add two new $P$s to $\mathcal{P}$ by forcing new constraints $x_i=1$ or $x_i=0$\;
    }
}
Optimal solution: $Z=c\hat{x}+b\hat{\nu}+\mathbb{E}[Q(\hat{x},\hat{\nu})]$
\caption{Branch-and-bound with dual decomposition}\label{Alg:MIP_BB}
\end{algorithm}

\section{Results and Discussion}\label{sec:Results}
\begin{figure}
    \centering
    \includegraphics[width = 0.8\columnwidth]{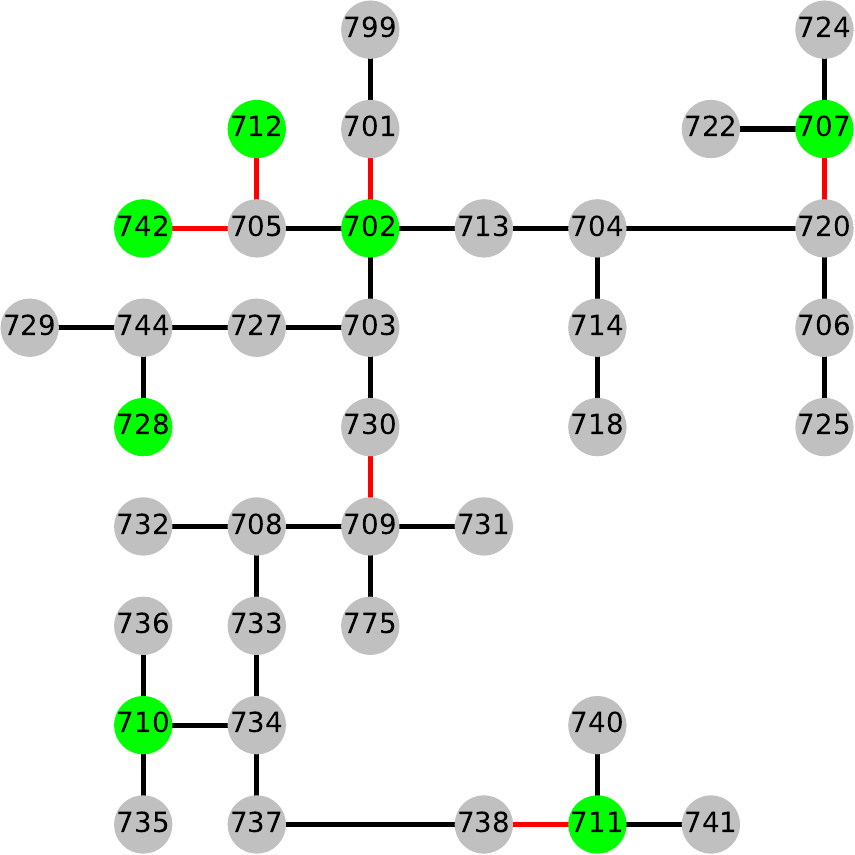}
    \caption{Layout of the IEEE 37-bus test feeder used in this study: red lines are the damaged edges, while green nodes are the candidate positions to host DERs}
    \label{fig:IEEE37}
\end{figure}

To validate the proposed recovery framework, the IEEE 37-bus test feeder is used as the ICI to restore after disruptions. And the disruption is set to be random line outages at 6 distribution lines, as red edges in \figureautorefname~\ref{fig:IEEE37}. The green nodes are the candidate nodes determined by the preassigning DERs step. Notice that two candidates $702$ and $728$ are chosen in the central disconnected component because of the scale of that disconnected subgraph after disruptions. As for the SO, 50 scenarios in total are sampled and the simulation time is set to be 24 time steps to represent 24 hours in a day. 2 repair modes are available for the solutions of $x_{ij}^r$, i.e. $r\in\{1,2\}$. And the normal distributed uncertain parameter $rs_{ij,r,s}$ is sample from $\mathcal{N}(\mu_{ij,r,s},\sigma^2)$ with two different levels of $\mu_{ij,r,s}$ at 5 and 10 units. On the other hand, $trepair_{ij,r,s}$ for each scenario is sampled based on \equationautorefname~\ref{eqn:MTTR}, with two different $\lambda_{ij,r,s}$ of 1 and 3 time steps. The MILP model is formulated and solved by Gurobi 8.1.1 on a server with dual E5-2650 v4 @ 2.20GHz CPUs and 128 GB memory.

\subsection{Comparing SO to Deterministic Formulation}
As discussed in previous sections, SO based method for decision making usually is more computational expensive than solving a deterministic problem. With the presence of uncertainties, stake holders may prefer to solve a deterministic version of the problem to achieve prompt actions. For example, the expected value of the random parameters, e.g. the repair time or required repair resource, can be taken into consideration to formulate a deterministic model. Hence, in order to demonstrate the effeteness of the PODIM framework, we first need to show the benefit of considering the randomness and formulate the problem as a two-stage SO.

This study utilizes two well-known measures, the expected value of perfect information (EVPI) and the value of stochastic solution (VSS) to quantify the effectiveness of the SO solutions. These two metrics can indicate the significance of adopting SO for solving the optimal solution. Nonetheless, these two metrics are defined based on the expectation solely, which means they can be only applied to the risk neutral model without any risk measure. In order to assess the solutions when SO problems involve with additional risk term, \citet{Noyan2012} extends the EVPI and VSS to be mean-risk value of perfect information (MRVPI) and mean-risk value of stochastic solution (MRVSS). To derive these two improved metrics for the risk-averse optimization, three different ways of formulating the SO need to be introduced first, as shown in \tableautorefname~\ref{tab:VSS}.
\begin{table*}[pos=t]
\def\arraystretch{1.5}
  \centering
  \caption{The definitions of three types of problem formulation to calculate the VPI and VSS of the risk-averse SO model as well as an exemplar numerical result from the PODIM}\label{tab:VSS}
    \begin{tabular}{cccc}
        \toprule
                          & Risk Neutral & Risk Averse (Mean Risk) & Result of PODIM with 50 scenarios\\
         \midrule
        Wait and See (WS) & $\mathbb{E}\left[\min\limits_{\mathbf{x}}f(\mathbf{x},Q)\right]$ & $\mathbb{E}\left[\min\limits_{\mathbf{x}}f(\mathbf{x},Q)\right]+\lambda\text{CVaR}_{\alpha}(\min\limits_{\mathbf{x}} f(\mathbf{x},Q))$ & 15557 \\
        Recourse Problem (RP) & $\min\limits_{\mathbf{x}}\mathbb{E}\left[f(\mathbf{x},Q)\right]$ & $\min\limits_{\mathbf{x}}\left\{\mathbb{E}\left[f(\mathbf{x},Q)\right]+\lambda \text{CVaR}_{\alpha}(f(\mathbf{x},Q))\right\}$ & 14927\\
        Expected Value (EV) & $\mathbb{E}\left[f(\bar{\mathbf{x}}(\bar{\mathbf{\xi}}),Q)\right]$ & $\mathbb{E}\left[f(\bar{\mathbf{x}}(\bar{\mathbf{\xi}}),Q)\right]+\lambda \text{CVaR}_{\alpha}(f(\bar{\mathbf{x}}(\bar{\mathbf{\xi}}),Q))$ & 13334\\

        \bottomrule
    \end{tabular}
\end{table*}

Here we represent the two-stage SO problem as $f(x,Q(\xi_s)),\\
\;\forall s\in \mathbf{S}$, where $Q$ again is the second stage problem and $\xi_s$ is one realization of the randomness. The wait-and-see (WS) program optimizes the problem for each scenario individually and reckons the overall expected objective value when perfect information presents. In other words, the WS approach assumes that complete information of the uncertainties is available before starting making decisions. The second program is the traditional deterministic equivalent formulation or the recourse problem (RP) for the scenario based SO. The RP is similar to the expression in \equationautorefname~\ref{eqn:SO}. And different from the WS, the RP does not utilize the perfect information of randomness and can only optimize the all-in-one problem in terms of expectation of all scenarios. The discrepancy between the WS and RP results defines the EVPI of the risk neutral optimization: $\text{EVPI}=\text{WS}-\text{RP}$. Besides, another formulation in \tableautorefname~\ref{tab:VSS} is the expected value (EV) problem. The EV first replaces the uncertainty $\xi_s$ in each scenario by the expectation of randomness $\bar{\xi}$. Then EV solves for the first stage decision variable based on this fixed uncertainty and we denote the solution as $\bar{x}(\bar{\xi})$. Correspondingly, the EV result is the expectation of objectives of all scenarios realized based on the presolved $\bar{x}(\bar{\xi})$. In this way, different uncertainties among scenarios are not taken into effect. And the VSS can be calculated as: $\text{VSS}=\text{RP}-\text{EV}$. On the basis of WS, RP, and EV for risk neutral problem, \citet{Noyan2012} introduces the mean-risk WS (MRWS), mean-risk RP (MRRP), and mean-risk EV (MREV) with the additional CVaR term. For all three metrics, the CVaR value with weight $\lambda$ is obtained at the same significance level $\alpha$ for calculating the VaR term. Furthermore, similar to the WS formulation, the MRWS derives the CVaR after the optimal result for each scenario has been solved individually. Whereas the MRRP finds the optimum of both the deterministic formulation and the CVaR measure for all scenarios in whole. And the MREV uses the expectation of uncertainties for calculating the CVaR. 

Once the MRRP, MRWS, and MREV are defined, by adopting the definition of EVPI and VSS, the MRVPI and MRVSS for risk-averse program can be calculated as:
\begin{gather}
\text{MRVPI} = \text{MRWS}-\text{MRRP},\\
\text{MRVSS} = \text{MRRP}-\text{MREV}.
\end{gather}
And to demonstrate the benefit of modeling the PODIM as a SO program, \tableautorefname~\ref{tab:VSS} summarizes the MRWS, MRRP, and MREV values of the PODIM problem for the IEEE 37-bus system after disruptions. All three metrics are measured in units of power restored in total. The corresponding MRVPI and MRVSS for the PODIM are found to be 630 and 1593 units of power restored correspondingly. In other words, based on the MRVSS, the solution has been improved by 11.95\% from the MREV formulation, after considering different uncertainties.
\begin{figure}
    \centering
    \includegraphics[width = 0.8\columnwidth]{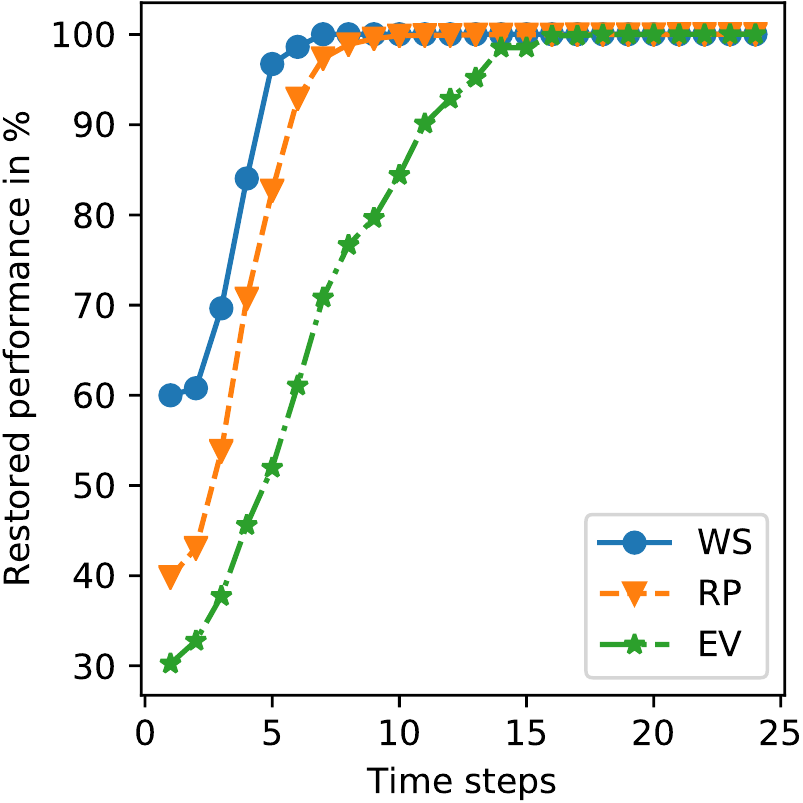}
    \caption{The average recovery profile of 50 scenarios for the IEEE 37-bus system governed by the PODIM with three formulations: WS, RP, EV ($\lambda = 1,\;\alpha=0.8$)}
    \label{fig:recovery_VSS}
\end{figure}

In addition to the numerical results, same discrepancies between WS, RP, and EV can be found in recovery profiles of the IEEE 37-bus system after disturbances. \figureautorefname~\ref{fig:recovery_VSS} shows the average system performance result from 50 scenarios, in which the performance during recovery is measured as the restored power in percentage. Based on the recovery profile, the SO approach of coordinating RCs and DERs in parallel has successfully restored the ICI to its nominal state (resilience level back to 1) for all three formulations of the PODIM. Nevertheless, the recovery profile for the MRWS is the best among the three formulations. Since uncertainties about repair time and required resource are deterministic and the optimal scheduling of DERs and RCs can be found for each scenario individually. But decision makers can hardly obtain MRWS solutions for practical applications because the perfect information of the future uncertainty is almost unavailable. On the other hand, the MREV performs the worst in terms of the resilience level. Since it naively considers the expectation of uncertainties of all scenarios as the true parameter. On the contrary, the performance curve of the MRRP formulation is just slightly worse than that of the ideal MRWS case. Thus, the MRRP approach is the most appropriate method to formulate the PODIM framework to enhance resilience.  

\subsection{Comparing CVaR to Risk-Neutral Formulation}
\begin{figure}
    \centering
    \includegraphics[width = 0.8\columnwidth]{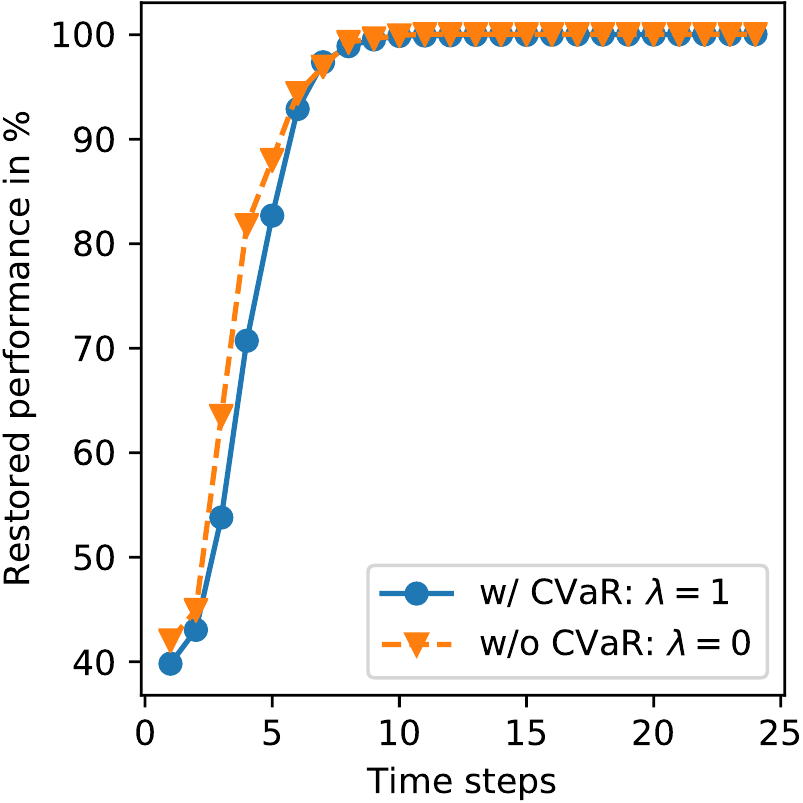}
    \caption{The average recovery profile for the IEEE 37-bus system governed by the PODIM with and without CVaR measure ($\alpha=0.8$)}
    \label{fig:recovery_CVaR}
\end{figure}

\begin{figure}
    \centering
    \includegraphics[width = 0.8\columnwidth]{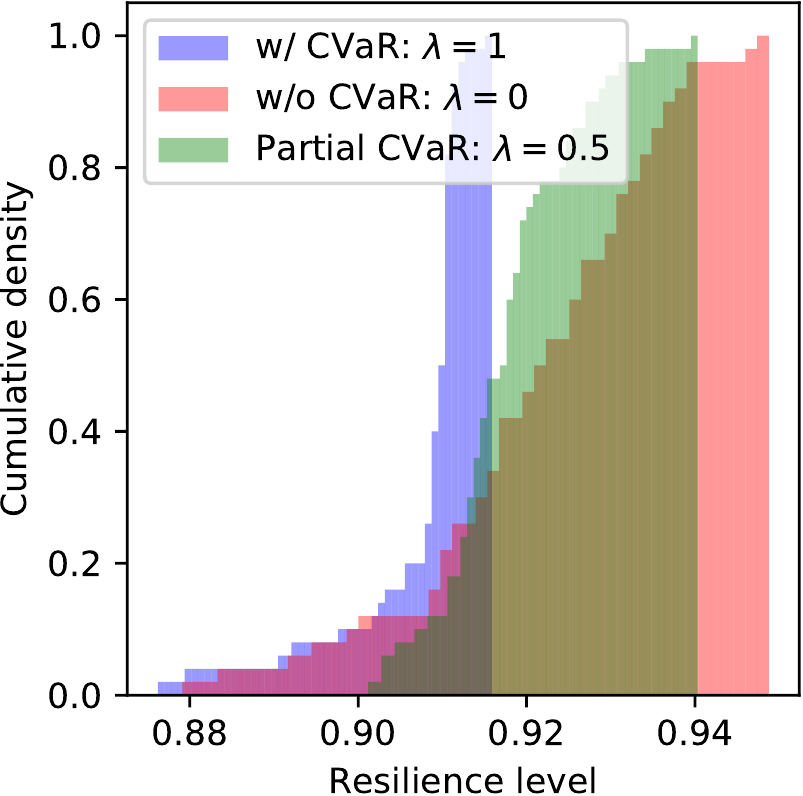}
    \caption{Cumulative density plot of the resilience level obtained in the solutions for 50 scenarios, by formulating PODIM in deterministic equivalent form with different weights for CVaR measure ($\alpha = 0.8$)}
    \label{fig:CDF_CVaR}
\end{figure}

To further validate the effects of adding CVaR into the PODIM, the average system performances under two scenarios, i.e. the PODIM with and without CVaR term, are illustrated in \figureautorefname~\ref{fig:recovery_CVaR}. 
In the plot, the average recovery profiles are very close to each other. The w/o CVaR case is even slightly better than the result having CVaR, as per the orange curve is above the blue curve at the first six time steps. Nonetheless, \figureautorefname~\ref{fig:recovery_CVaR} displays the system recovery performance in terms of the expectation of all solutions for 50 scenarios. It's known that including additional risk term reduces the variance of the random solutions but slightly deteriorates the random outcomes in expectation. In order to manifest the trade-off between the variance and expectation, \figureautorefname~\ref{fig:CDF_CVaR} summarizes the cumulative density of resilience level results calculated by following \equationautorefname~\ref{eqn:resilience} for 50 scenarios. Each histogram corresponds to one setting for the weight of CVaR in the PODIM model.

Comparing the results of incorporating CVaR in the objective (blue) to that of conventional deterministic equivalent formulations (red) without CVaR, we can see that the additional risk measure helps the SO solutions attain much less variance, which is consistent with the benefit of applying the risks averse optimization. However, we can see that the expectation of the resilience level of the solutions obtained from the risk-averse method is smaller than that of deterministic equivalent formulation, which has been seen in \figureautorefname~\ref{fig:recovery_CVaR}. To accommodate this trade-off between the expectation and variance of the SO solutions, the weight of the risk measure, i.e. $\lambda$ in \equationautorefname~\ref{eqn:Objective} can be adjusted to an appropriate level. And one example is given as the green histogram in \figureautorefname~\ref{fig:CDF_CVaR}. It can be seen that solutions with $\lambda=0.5$ not only have smaller variance than solutions without CVaR, but also acquire higher expectation than the resilience levels with $\lambda=1$.
\begin{figure*}
\centering
\begin{subfigure}{0.4\textwidth}
  \centering
  \includegraphics[width=\linewidth]{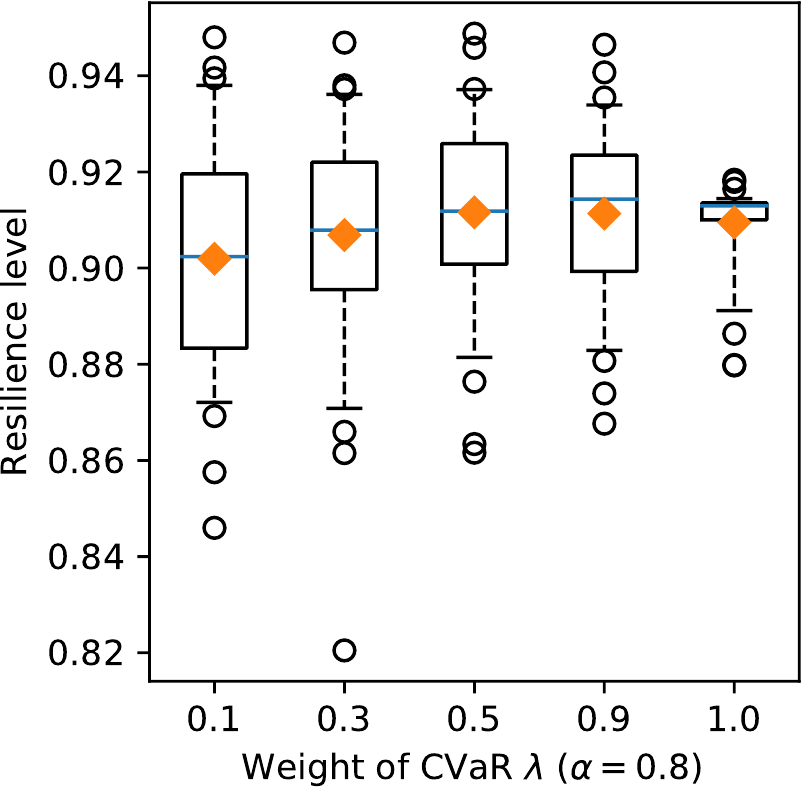}
  \caption{The resilience level versus the weight $\lambda$ of the CVaR risk measure ($\alpha$ fixed at 0.8)}
  \label{fig:box_1}
\end{subfigure}
\qquad
\begin{subfigure}{0.4\textwidth}
  \centering
  \includegraphics[width=\linewidth]{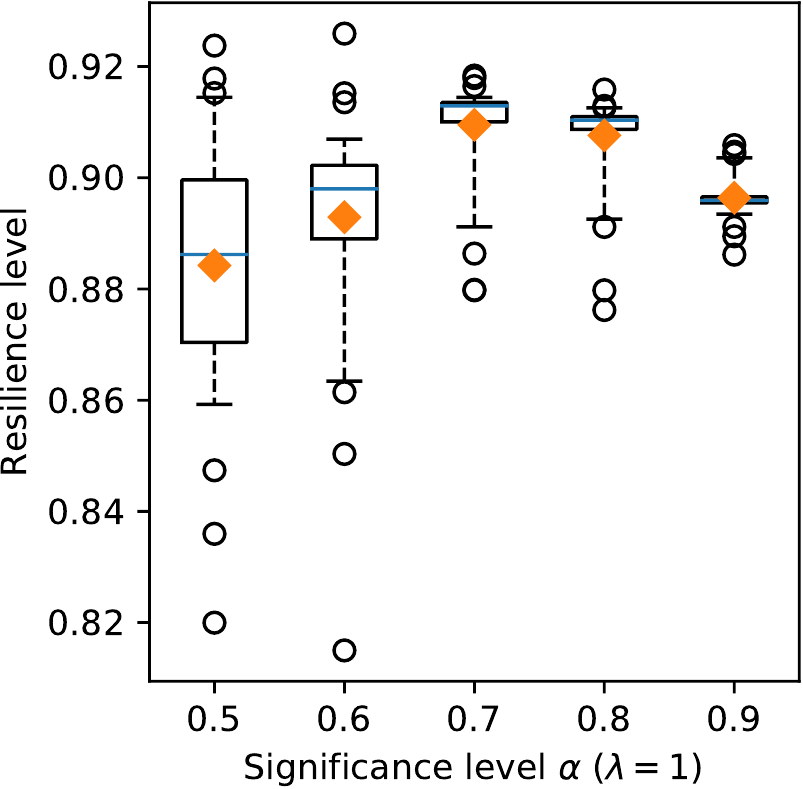}
  \caption{The resilience level versus the significance level $\alpha$ in the CVaR formulation ($\lambda$ fixed at 1)}
  \label{fig:box_2}
\end{subfigure}
\caption{Box plots show the sensitivity analysis results for the resilience level of the IEEE 37-bus system}
\label{fig:box}
\end{figure*}

Besides the weight of the risk measure $\lambda$, other hyperparameters such as the significance level $\alpha$ can also change the final resilience results. In order to comprehend the effects of several hyperparameters of the model on the recovery performance, sensitivity analysis about $\lambda$ and $\alpha$ is conducted, and the results are summarized as box plots in \figureautorefname~\ref{fig:box}. The blue line in each box indicates the median level of the solutions while the orange diamond represents the expectation. Based on \figureautorefname~\ref{fig:box_1}, we can see that increasing the weight of the risk measure from 0.1 to 1.0 slightly changes the mean of the resilience level solutions from 0.902 to 0.909, but the solutions become much more concentrated with respect to the variance. This is because the SO problem puts more efforts on minimizing the additional risk and thus the number of outliers shrinks significantly, as shown in \figureautorefname~\ref{fig:CDF_CVaR}. And \figureautorefname~\ref{fig:box_2} indicates that a significance level around 0.7 to 0.8 leads to the best result with the highest expectation as well as relatively low variance. To further provide readers with a clear picture of how the hyperparameters will change the risk-averse solutions, numerical results from the IEEE 37-bus system with the proposed PODIM under four different settings are outlined in \tableautorefname~\ref{tab:sensitivity}. The results in the table shows the same trade-off between the expectation and variance. Additionally, results of the case when $\lambda=1,\alpha=0.5$ suggest that poor setting of the hyperparameters can even deteriorate the performance of the SO based PODIM.
\begin{table*}[pos=t]
    \centering
    \caption{Numerical results for the sensitivity analysis in terms of the resilience level achieved}\label{tab:sensitivity}
    \begin{tabular}{ccccccc}
    \toprule
        PODIM Setting & Min & $25\%$ Quantile & Mean & $75\%$ Quantile & Max & Variance \\
        \midrule
    $\lambda=0$ & 0.8791& 0.9125& 0.9211& 0.9317& 0.9488& 2.198e-4\\
    $\lambda=1$, $\alpha=0.5$ & 0.8200& 0.8704& 0.8842& 0.8996& 0.9238& 4.480e-4\\
    $\lambda=1$, $\alpha=0.9$ & 0.8862& 0.8955& 0.8964& 0.8965& 0.9059& 1.018e-5\\
    $\lambda=0.1$, $\alpha=0.8$ & 0.8460 & 0.8833& 0.9019& 0.9196& 0.9480& 5.392e-4\\
    $\lambda=1$, $\alpha=0.8$ & 0.8798& 0.9100& 0.9095& 0.9135& 0.9184& 7.449e-5\\
    \bottomrule
    \end{tabular}
\end{table*}

\subsection{Time Complexity}
Lastly but not the least, to point out the advantage of incorporating the Lagrangian dual based decomposition technique for solving the PODIM, \tableautorefname~\ref{tab:time} shows the solving time of the MILP model with and without the DD process for two different settings. The incumbent is the best feasible solution found by the branch-and-bound algorithm. And the best bound is the optimal objective value of the relaxed problem, in which solutions are still not feasible but are better than the current feasible solutions (incumbent). The mixed integer programming (MIP) gap is defined as the difference between the feasible solution (incumbent) and the relaxed solution (best bound) in percentage. 
\begin{table*}
    \centering
    \caption{Comparison of the computational performance for different settings}\label{tab:time}
    \begin{tabular}{cccccc}
    \toprule
         Setting & Decomposition & Time & Incumbent & Best Bound & MIP Gap\\
         \midrule
         \multirow{2}{*}{$\lambda=0$} & No Decomposition & 10111s& 7018& 7453& 5.84\%\\
         \cmidrule(lr){2-6}
         {} & With Decomposition & 8281s& 7278& 7506 & 3.04\%\\
         \midrule
         \multirow{2}{*}{$\lambda=1,\;\alpha=0.8$} & No Decomposition & 10758s& 7071& 7352& 3.82\%\\
         \cmidrule(lr){2-6}
         {} & With Decomposition & 7648s& 7077& 7356 & 3.79\%\\
    \bottomrule
    \end{tabular}
\end{table*}
Form the table, for both the risk neutral ($\lambda=0$) and the risk-averse ($\lambda=1$) case, the computational time for obtaining incumbents around the same level has been reduced by more than 25\% after utilizing the decomposition technique. Besides, the model with the decomposition achieves half of the MIP Gap within shorter time for the risk neutral case. And the risk-averse optimization with DD reaches the same MIP gap ($\approx 3.8\%$) one hour earlier comparing to the case without DD. Notice that for both cases, the MIP gap is not 0\% even after several hours. This is because mathematical programming problems involving with integer variables usually are NP-hard. Finding the global/true optimum of a complex MILP model, such as the PODIM framework proposed in this paper, can take extremely long time. Even though the global optimal solution is hard to find, the MIP gap helps decision makers evaluate the quality of the feasible solutions obtained and stop the solving process in advance.
\begin{figure*}
\centering
\begin{subfigure}{0.4\textwidth}
  \centering
  \includegraphics[width=\linewidth]{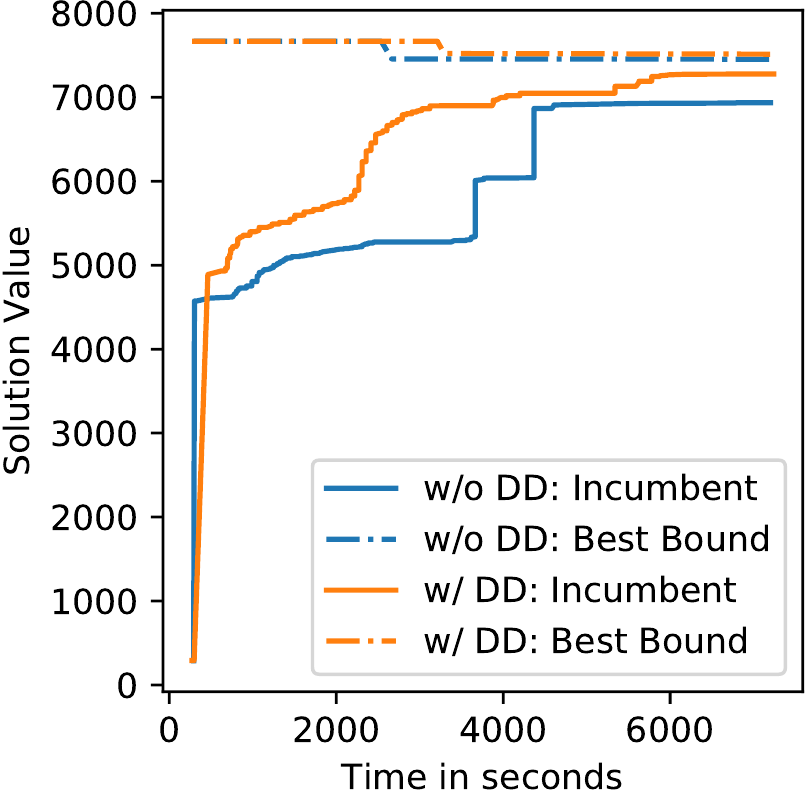}
  \caption{The incumbent and best bound solved for the DE formulation with $\lambda=0$, for both with and without dual decomposition cases}
  \label{fig:converge_1}
\end{subfigure}
\qquad
\begin{subfigure}{0.4\textwidth}
  \centering
  \includegraphics[width=\linewidth]{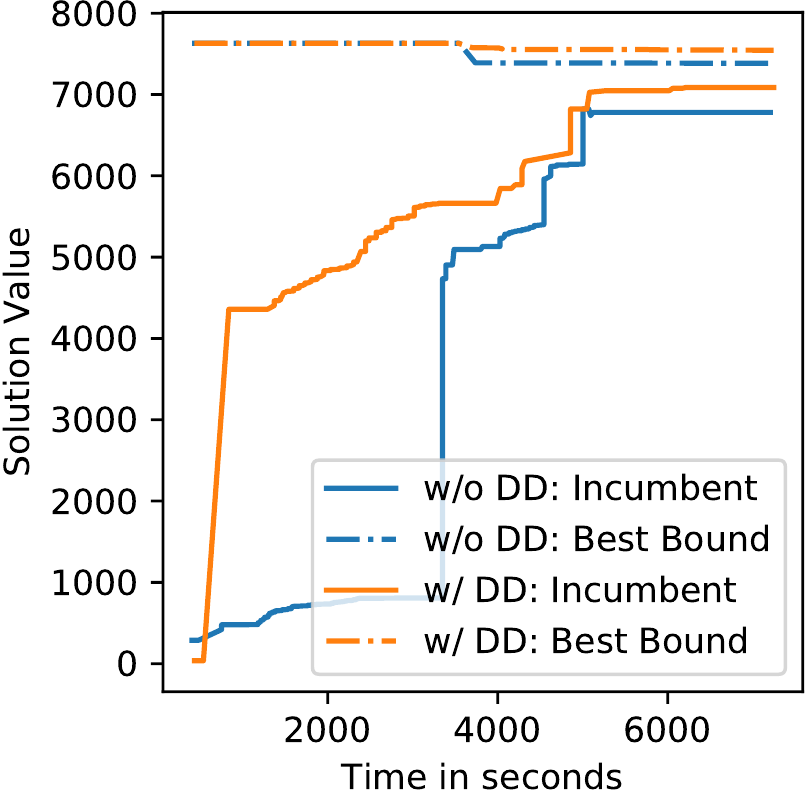}
  \caption{The incumbent and best bound solved for the DE formulation with $\lambda=1$, for both with and without dual decomposition cases}
  \label{fig:converge_2}
\end{subfigure}
\caption{The convergence plot of the solving process during the first two hours}
\label{fig:converge}
\end{figure*}

In addition to the numerical results of the solving process, \figureautorefname~\ref{fig:converge} shows the convergence plot of the solutions. Here only results of the first two hours are presented. For both the cases when setting the $\lambda$ to be 0 and 1, the model with DD comes by a smaller gap between the incumbent and the best bound than the model without DD in the same time horizon. Moreover, we can see the DD helps accelerate the convergence significantly between the incumbent and best bound, especially during early steps before 4000 seconds for both risk neutral and risk-averse cases. This fast convergence behavior at initial steps motivates a future direction of solving the two-stage SO based PODIM:
\begin{enumerate}
    \item predefine a MIP gap for stopping the brach-and-bound algorithm;
    \item solve the MILP model by the DD introduced in \sectionautorefname~\ref{sec:Solving}; stop when reach the MIP gap;
    \item based on the suboptimal feasible solutions, use meta heuristic approaches, e.g. Tabu search or genetic algorithm, to converge to the true optimum.
\end{enumerate}
In this way, in stead of keeping running the branch-and-bound algorithm with exponentially growing search tree, the heuristic algorithm could approach the global optimum surprisingly faster due to its inherent randomness.

\section{Conclusion}\label{sec:Conclusion}
This work presents a PODIM framework for ICI systems after disruptions in order to ensure the resilience. Coordination of two recovery agents, RC and DER, is realized by modeling the recovery process as a MILP problem. And in order to tackle the uncertainties presenting in the recovery process, e.g. random repair time and required resource for repairing, the framework is extended to be a two-stage SO problem using the deterministic equivalent form. Furthermore, this study adopts the risk-averse optimization technique to address the challenge of temporal sparsity inherented in the PODIM problem for ICIs. Case study results based on the IEEE 37-bus test feeder demonstrate benefits of using the developed risk-averse optimization framework for resilience improvement as well as the advantages of adopting SO formulations in the restoration planning for ICIs.

\section*{Acknowledgement}
This research is partially supported by the National Science Foundation through the Faculty Early Career Development (CAREER) award (CMMI-1813111), the Engineering Research Center for Power Optimization of Electro-Thermal Systems (POETS) with cooperative agreement EEC-1449548, and the U.S. Department of Energy's Office of Nuclear Energy under Award No. DE-NE0008899.

\printcredits

\bibliographystyle{model1-num-names}
\bibliography{mybibfile}
\end{document}